# Photo-induced blinking in a solid state quantum system

Amanuel M. Berhane[1], Carlo Bradac[1] and Igor Aharonovich[1,*]

[1] School of Mathematical and Physical Sciences, University of Technology Sydney, Ultimo, New South Wales 2007, Australia.

*igor.aharonovich@uts.edu.au

*Solid state single photon emitters (SPEs) are one of the prime components of many quantum nanophotonics devices. In this work, we report on an unusual, photo-induced blinking phenomenon of SPEs in gallium nitride (GaN). This is shown to be due to the modification in the transition kinetics of the emitter, via the introduction of additional laser-activated states. We investigate and characterize the blinking effect on the brightness of the source and the statistics of the emitted photons. Combining second-order correlation and fluorescence trajectory measurements, we determine the photo-dynamics of the trap states and characterize power dependent decay rates and characteristic "off"-time blinking. Our work sheds new light into understanding solid-state quantum system dynamics and, specifically, power-induced blinking phenomena in SPEs.*

Fluorescence blinking, also referred as fluorescence intermittency, is usually an undesired but ubiquitous phenomenon in most quantum light sources, including quantum dots (QDs) [1-3], defects in wide-bandgap semiconductors [4-6] and single molecules[7-9]. Blinking arises when, upon laser excitation, a fluorescent centre undergoes sporadic jumps between "dark" and "bright" states in the photo-emission [3]. This phenomenon is identified by the random fall ("off"/"dark" state) and rise ("on"/"bright" state) in photon counts during long time (milliseconds to hours) fluorescence photostability measurements. Although the cause of the "dark" state has been rigorously studied in various fluorescent systems, a universal physical mechanism that explains blinking has not been pinned down yet [2,10-14]. In most cases, blinking results in photo-bleaching of the emitter, in the permanent "off" state.

Recently, a new family of single-photon emitters (SPEs) in Gallium Nitride (GaN) has been discovered [15]. Using both experimental and modelling techniques, the single photon emission was attributed to the recombination of localised excitons to a point defect sitting near or inside a cubic inclusion. These emitters show bright, narrow-band emission with linear polarization, which is suitable for quantum information applications. Under continuous wave laser excitation, the vast majority of the emitters display photo-stable fluorescent emission with near-Poissonian statistics. Interestingly however, approximately 5% of the emitters start showing blinking once the power of the excitation laser rises over a certain threshold.

In this work, we investigate the nature of this excitation-induced blinking behaviour of SPEs in GaN, at room temperature. By combining transition kinetics analysis and fluorescence correlation measurements at short (nanoseconds) and long (millisecond) time scales, we gather new insights into the blinking mechanism. Furthermore, we propose a mechanism to explain this behaviour in the attempt to generalize the phenomenon and extend the description of such laser-induced blinking to other solid state SPEs.

The sample used in this study is a 2-µm thick Magnesium (Mg)-doped GaN layer on 2-µm undoped GaN grown on sapphire which is commercially available (Suzhou Nanowin Science and Technology Co., Ltd.). The SPEs were isolated at Room-Temperature (RT) using a custom-made confocal microscope [15] equipped with a Hanbury-Brown and Twiss (HBT) interferometer for second-order autocorrelation measurements. A 532-nm, cw laser was used for excitation and the laser power was measured at the back-focal plane of the confocal microscope objective.

Figure 1a shows the photoluminescence spectrum of an isolated SPE excited with 200 µW of laser power, at room temperature. The emitter displays a characteristic emission with zero-phonon line (ZPL) at 647 nm and Full Width at Half Maximum (FWHM) of ~4 nm. Figure 1b shows the emitter's fluorescence stability, measured at 3 mW. The corresponding occurrence statistics of the emission intensity is shown in Figure 1c. The photon distribution follows a near-Poissonian statistics at excitation power of 3 mW[16].

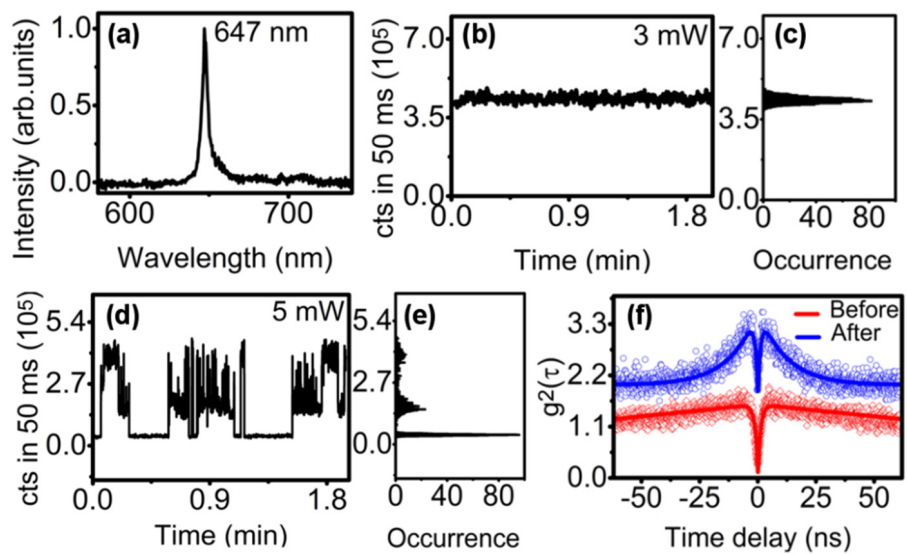

**Figure 1**. Excitation power-induced blinking of a SPE in GaN. **a)** RT spectra of the SPE taken under 200-µW power excitation; the ZPL lies at 647 nm with a FWHM of ~4 nm. **b)** Fluorescence photostability and **c)** occurrence statistics of the emitter (data taken at 3 mW over 2 minutes). The emitter shows stable emission with near-Poissonian photon statistics. **d)** Fluorescence photostability and **e)** photon occurrence statistics of the same emitter excited at the higher excitation power of 5 mW: at this power blinking appears (time binning in all measurements is 50 ms). **f)** Autocorrelation function $g^2(\tau=0)$ measured for the same emitter before (red), and after blinking (blue) was induced with high-power excitation; the $g^2(0)$ curves in (f) are both taken with 100 µW excitation power, the [blue] curve is offset for displaying purposes. The emitter clearly shows a permanent change in behaviour with a lower contrast in $g^2(0)$ after the high-power-excitation blinking was induced: $g^2(0) \approx 0.65$ vs. $g^2(0) \approx 0.24$ before and after, respectively.

For this particular centre, a 5-mW excitation induced a sudden change in the photon statistics, which starts displaying a marked blinking behaviour as illustrated in Figure 1d. The photon statistics of the emitter at 5 mW (Fig. 1d, e), shows three distinct states, in contrast to the near-Poissonian single state of the same emitter at 3 mW (Fig. 1b, c). We note that this blinking did not result in bleaching of the emitter during the time frame of this characterization. The fluorescence trajectories – before and after the blinking is induced – for different excitation intensities are shown in Figure S2. The average intensity values of the emitter before and after blinking are given in Table S1.

Prompted by the unusual blinking characteristics of the SPEs, second-order autocorrelation $g^2(\tau)$ measurements were taken at different excitation powers before and after blinking, using a Hanbury-Brown and Twiss (HBT) interferometer. Figure 1f shows two $g^2(\tau)$ curves of the same SPE taken at excitation power of 100 µW before (red) and after (blue) blinking was induced. A sequence diagram is provided in Figure S1 to illustrate how the measurement was carried out. Remarkably, the photon statistics before and after the induced blinking are different, with the emitter showing – beyond the expected bunching at intermediate time scales due to the high excitation power – a reduction in the contrast of the $g^2(0)$ function ($g^2(0) = 0.65$ after, vs. 0.24 before). This indicates that the transition dynamics of the emitter is permanently modified. We argue that the change is caused by the activation of a trap state which provides an additional, non-radiative transition pathway to the ground state, before the system can be re-excited (level diagram is shown in Figure S3). Note that this behaviour is dramatically different from that of other solid state emitters – e.g. the nitrogen vacancy (NV) centre in diamond – where high excitation simply results in an increased population of its metastable state, and the photo-dynamics is preserved [17-19].

The background-corrected, second order autocorrelation curves in Figure 1f are fitted with a three-level model given in equation (1).

$$g^2(\tau) = 1 - (1+a)e^{-[\lambda_1 \tau]} + ae^{-[\lambda_2 \tau]} \qquad (1)$$

Where $\lambda_1$ and $\lambda_2$ are fitting parameters for radiative and non-radiative decay rates while $a$ is a scaling factor for bunching. With an excitation power of 100 µW, $g^2(0) = 0.24$ and 0.65 'before' and 'after' blinking, respectively. While the contrast in $g^2(0)$ is reduced after the power-induced blinking has occurred we believe it still being associated with the same single emitter, only with a much lower signal to noise ratio (i.e. reduced brightness as per the additional dark shelving state). Additional power-dependent autocorrelation $g^2(0)$ curves before and after blinking are shown in Figure S4.

To further pin down the changes in transition kinetics due to the laser-induced trap state, we conduct an analysis of the emitter's brightness and rate coefficient before and after blinking. Figure 2a shows plots of power-dependent intensity values for the SPE before and after the blinking was induced. The plots are fitted with a three-level system model, and the intensity is given by:

$$I = I_\infty \frac{P}{P + P_{sat}} \qquad (2)$$

$P_{sat}$ is the saturation power and $I_\infty$ is the highest intensity obtained. Before blinking, $I_\infty$ is ~527 kcounts/s at $P_{sat}$ ~660 µW. The same centre showed two different saturation behaviours before and after blinking. After the blinking is induced, the emission intensity at excitation powers <900 µW is slightly lower than it was before blinking (for the same powers). This is consistent with the model we propose of a laser-activated trap state compounding the non-radiative transition. At excitation powers ≥900 µW, the effect of the additional trap state on emission intensity is overall reduced due to rapid depopulation to the ground state [20,21].

The transition kinetics analysis is carried out by extracting $\lambda_1$, $\lambda_2$ and $a$ as fit parameters from the background-corrected, power-dependent second-order correlation measurements before and after blinking as shown in Figures S4a and b. Figures 2b-d display the extracted $\lambda_1$, $\lambda_2$ and $a$ as a function of excitation powers, before and after blinking was induced. The power dependence of $\lambda_1$, $\lambda_2$ and $a$ is fitted by assuming a three-level model with a shelving state that depends linearly on the excitation power for both 'before' and 'after' blinking (detailed analysis is presented in the SI) [22-24]. The parameter $\lambda_1$ before and after blinking remains unchanged, hinting that the radiative decay pathway is unaffected by the blinking. Note that the ZPL in the spectra remains unchanged after blinking was induced, suggesting that whatever the nature of the laser-induced change in the emitter might be, such change does not alter, in a detectable manner, the first excited electronic state. Conversely, there is a clear difference in the distribution of values for $\lambda_2$ and $a$, where the three-level model for the emitter fails to fit the power-dependent behaviours of $\lambda_2$ and $a$ after blinking. This is also in accord with the pronounced bunching behaviour at intermediate time scales observed for the emitter at low powers after the induced blinking.

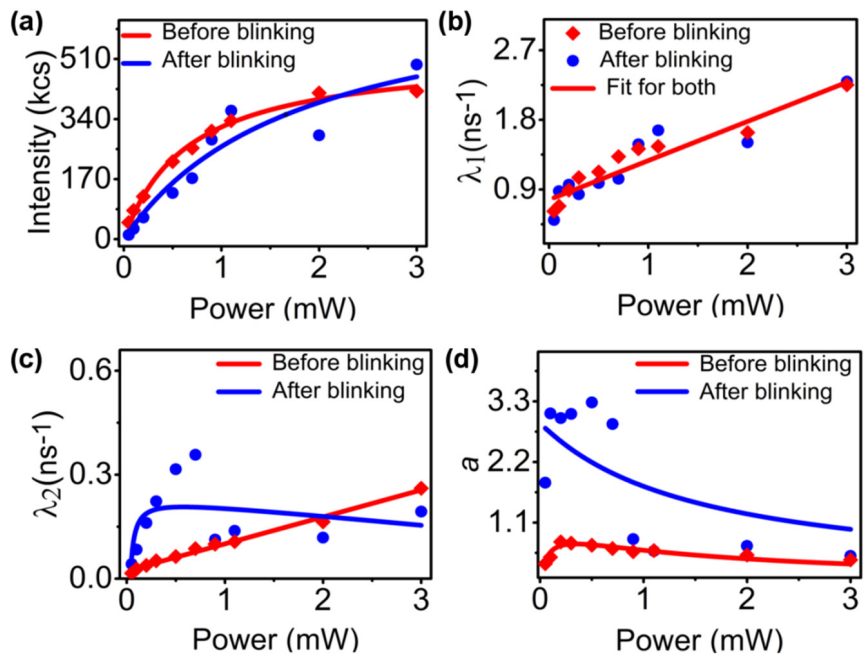

**Figure 2.** Excitation-power dependent parameters of the emitter. **a)** Brightness of the emitter before (red) and after (blue) blinking. Before blinking, at saturation power ($P_{sat}$) ~660 µW, the highest intensity of 527 kcounts/s is obtained. After blinking, the saturation behaviour is fitted with a three-level model showing a remarkably different curve. **b–d)** Power-dependent characteristics for the fit parameters $\lambda_1$, $\lambda_2$, $a$, respectively, for the $g^2(\tau)$ function. These values are extracted as parameters from the $g^2(\tau)$ function fitting (SI, Fig. S4). A three-level model with linear power dependence for the shelving state described accurately the transition kinetics before blinking (red fitting lines). After blinking, however, the same model fails to fit $\lambda_2$ and $a$ as highlighted by the blue lines in (c) and (d).

While rate analysis and brightness characterization hint to a permanent change in the photo-dynamics of the emitter, a more direct evidence for the power-induced trap state is required. We therefore recorded long time-fluorescence correlation behaviours for two different emitters in the time range of a few microseconds to 0.1 seconds – one that exhibits absolute photostability and another one that exhibits blinking at higher excitation powers (similar to the one characterized earlier). The spectra and saturation behaviours of the two SPEs is shown in Figure S5. Figure 3a shows power-dependent, long-time-scale correlated $g^2(\tau)$ from the photostable SPE. Each measurement is fitted with exponential decay function that hold the least chi-square value, where the decay rate determines the bunching behaviour [25]. In this case, no significant (compared to noise level) decay rates can be observed at the microsecond-to-millisecond time scale and the $g^2(\tau)$ remains constant along the normal line for all excitation powers. This suggests that for the stable emitter, the corresponding photon statistics at different excitation powers shows no sign of blinking at longer time scales.

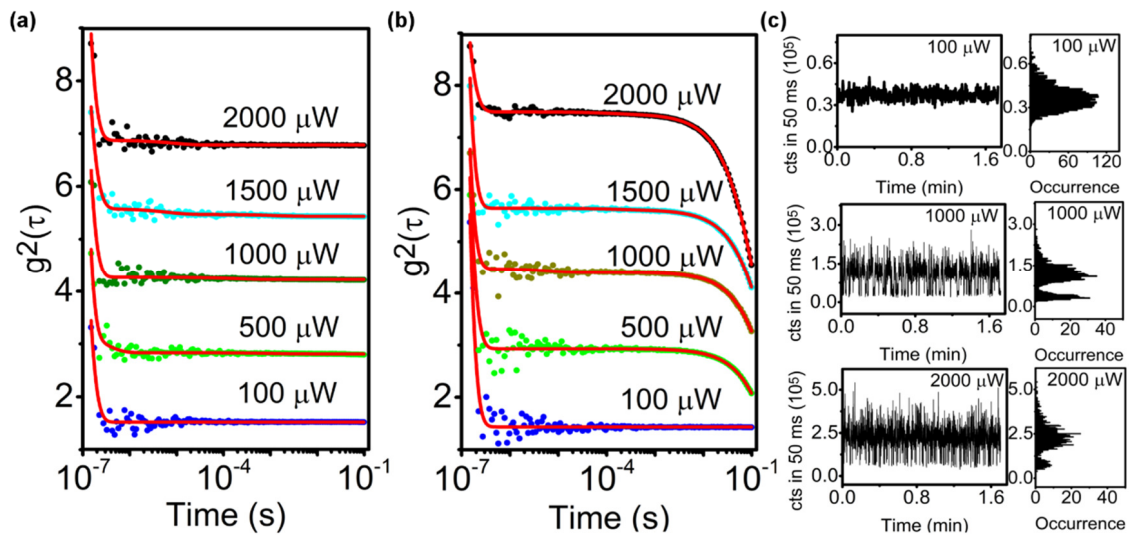

**Figure 3.** Power-dependent long lifetime fluorescence of a stable and a blinking emitter. **a)** Long time scale, excitation-power-dependent $g^2(\tau)$ characteristics of a stable emitter. The best fit is determined using a single and double exponential decay function with the least chi-square value. $g^2(\tau)$ starts with monotonic decay that corresponds to ns shelving state but remains constant for the measurement time scale range of microseconds to 0.1 seconds. **b)** Long time scale $g^2(\tau)$ characteristics of a power-induced blinking emitter at different excitation powers. Fitting the $g^2(\tau)$ characteristics at excitation powers of 100 µW is done using single exponential decay function where $g^2(\tau)$ remained constant along the normal; whereas, for excitation power of 500–2000 µW, the emitter showed an additional bunching curve in the ms range with the height increasing with power. **c)** Fluorescence photostability and photon occurrence statistics under 50-ms binning for the emitter in (b) with increasing excitation power. The emitter, initially stable with near-Poissonian statistics, starts blinking for excitation powers ≥500 µW.

The same set of measurement and analysis is carried out for an emitter that exhibited excitation-dependent blinking. In this case, $g^2(\tau)$ measurement at long time scales show strong dependence on the excitation power as evident from Figure 3b, where an extra decay channel appears at the ms scale. A qualitative difference can also be spotted with increasing power. At 100 µW, $g^2(\tau)$ remains constant along the normal at the ms scale – much like the stable emitter at all excitation powers. At higher excitation powers, starting from 500 µW, $g^2(\tau)$ shows additional bunching decay in the ms range [16]. This is a direct evidence for a power-induced change in the emitters' photodynamics. By fitting $g^2(\tau)$ with double decay functions for excitation power in the range 500–2000 µW, we determined characteristic decay times for the longer decay channel of 371 ms, 222.4 ms, 181.8 ms and 110.4 ms at 500 µW, 1000 µW, 1500 µW and 2000 µW, respectively. The decay time of the induced trap state decreases with increasing excitation power, showing rapid depopulation from the newly activated trap state at higher excitation powers. This is consistent with the observed reduction of blinking events with increasing power and similar intensity after blinking was induced (as per in Fig. 2a).

Figure 3c shows the fluorescence photostability analysis corresponding to the blinking emitter displayed in Figure 3b: binning time is 50 ms and excitation powers are 100 µW, 1000 µW and 2000 µW. As per before, to the right of each photostability time trace, the corresponding photon statistics is displayed, with 1000 photon binning. At 100 µW, the emitter showed stable emission with the intensity trace displaying no "dark" state interruptions as well as minimal photon statistics deviation from near-Poissonian. Upon further increase of the excitation power from 500 µW to 2000 µW, long time scale blinking set in with the fluorescence photostability showing clear "on" and "off" times. The alternating "on" and "off" events are observed on the photon statistics, with the occurrence of the "off" states decreasing with increasing excitation power.

To quantify the "on" and "off" times at different excitation powers, the probability density distribution $P[\tau_{on}]$ and $P[\tau_{off}]$ were plotted by setting a threshold intensity on the fluorescence trajectories of the blinking (Fig. 3c) [26]. In the fluorescence time-trace, above and below the set threshold the emitter is considered to be "on" ($\tau_{on}$) or rather "off" ($\tau_{off}$), respectively.

$$P(\tau_{on}, \tau_{off}) \approx \exp\left(\frac{-\tau}{(\tau_{on}, \tau_{off})}\right) \qquad (3)$$

This relation shows a linear distribution on log-linear plots as displayed in Figure 4 for different excitation powers. Unlike the widely-reported power-law dependence of the probability density distribution on the $\tau_{on}$ and $\tau_{off}$ [26-29], single exponential decay of the form shown in Equation 3 is observed for both "on" and "off" times in the emitters we analysed [6,30]. From these plots, the characteristic decay times for $\tau_{on}$ and $\tau_{off}$ are determined for the excitation-dependent blinking, discussed above.

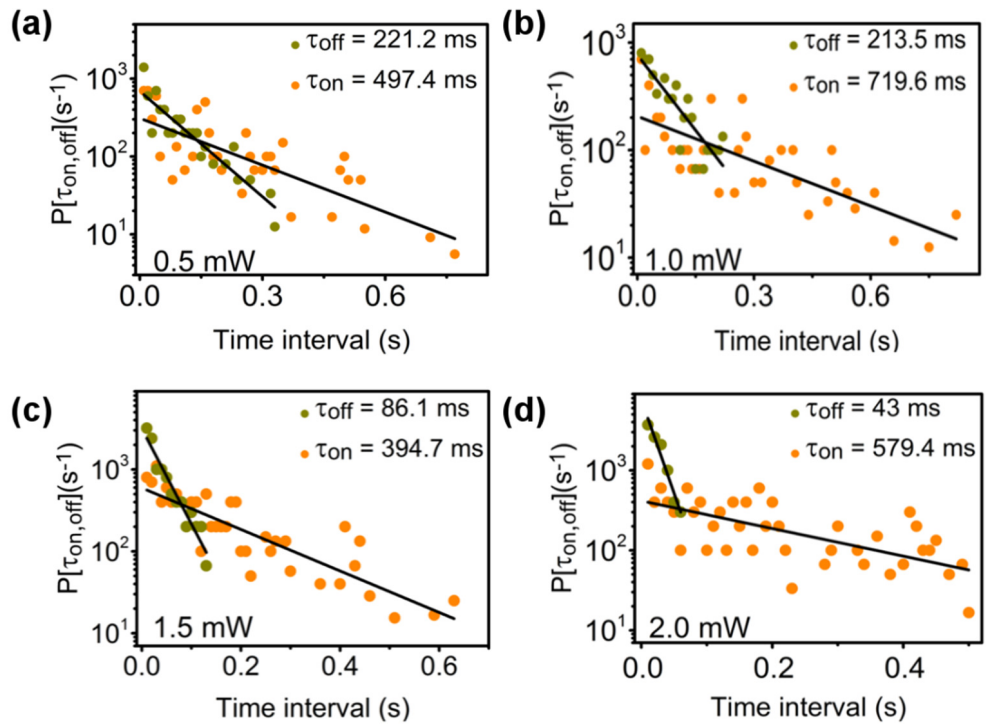

**Figure 4.** Probability distribution of "on" (orange) and "off" (dark yellow) states of the blinking emitter at different excitation powers. **a–d)** Semi-log plots of the "on" and "off" time distributions of the fluorescence trajectories shown in Figure 3c. The probability distributions of both the "on" and "off" times at all excitation powers show exponential decay, as indicated by the linear trend on the semi-log plots. The "off" probability

distributions hold characteristic decay time ($\tau_{off}$) that drop with increasing excitation power starting at 221.2 ms, 213.5 ms, 86.1 ms and 43.1 ms for power excitation in the range 50–2000 µW. Conversely the "on" time distribution did not show dependence on excitation power and gave a mean characteristic decay time ($\tau_{on}$) of (548±137) ms.

For excitation powers in the range 500–2000 µW, characteristic "off"-time, $\tau_{off}$, were measured to be ~221.2 ms, 213.5 ms, 86.1 ms and 43.1 ms, respectively. The decline in the "off" times with increasing power is consistent with the fitting at long time scale of $g^2(\tau)$ (Fig. 3b) and with the reduced blinking at high powers due to faster depopulation from the activated trap state.

Conversely, the probability distribution of "on" times did not show significant dependence on the excitation power, yielding a mean "on" time ($\tau_{on}$) of (548±137) ms. The weak dependence was also observed on the florescence statistics, where the "on" event occurrence remained the same while the "off" event decreased. This suggests that the depopulation of the new trap state possibly leads to a population of another shelving state. This is indicated by the rate coefficient, $\kappa_{43}$, in Figure S3. However, because the original shelving state and the induced trap state are weakly coupled in time, as demonstrated already via the second order autocorrelation function analysis of the transition kinetics, the effective increase in "on" times is not significant.

In conclusion, we presented a comprehensive investigation of excitation, power-dependent blinking of SPEs in GaN. We demonstrated that the excitation power permanently activates trapping states which act as additional shelving states associated with blinking. This is in contrast to known emitters in solids that, upon higher excitation, populate the same shelving/metastable state, or, quantum dots and single molecules that tend to bleach shortly after blinking. Our work has also practical implications. For example, the fact that the "on" time state remains relatively unchanged, means that the emitters in GaN can be used in practical quantum photonic applications (e.g. two-photon interference) where high photon flux is important [31,32]. Overall, our work helps shedding more light onto a rather complicated phenomenon – blinking in solid state SPEs – and emphasizes that standard three-level models may not always be ideal to describe the photo-dynamics of such systems.

**Acknowledgments**

Financial support from the Australian Research Council (via DE130100592), FEI Company, and the Asian Office of Aerospace Research and Development grant FA2386-15-1-4044 are gratefully acknowledged. The authors thank Milos Toth and Bernd Sontheimer for useful discussions.